\documentclass[prl,twocolumn,epsfig,rotate,superscriptaddress]{revtex4}
\usepackage{graphicx}
\usepackage{amsmath}
\usepackage{dcolumn}
\usepackage{epsfig}
\usepackage{bm}
\usepackage{mathrsfs}
\begin{document}

\title{One-dimensional Superdiffusive Heat Propagation Induced by Optical Phonon-Phonon Interactions}

\author{Daxing Xiong}
\affiliation{Department of Physics, Fuzhou University, Fuzhou 350108, Fujian, China}

\author{Yong Zhang}
\email{yzhang75@xmu.edu.cn}
\affiliation{Department of Physics and Institute of Theoretical Physics and Astrophysics, Xiamen University, Xiamen 361005, Fujian, China}

\begin{abstract}
It is known that one-dimensional anomalous heat propagation is usually characterized by a L\'{e}vy walk superdiffusive spreading function with two side peaks located on the fronts due to the finite velocity of acoustic phonons, and in the case when the acoustic phonons vanish, e.g., due to the phonon-lattice interactions such that the system's momentum is not conserved, the side peaks will disappear and a normal Gaussian diffusive heat propagating behavior will be observed. Here we show that there exists another type of superdiffusive, non-Gaussian heat propagation but without side peaks in a typical nonacoustic, momentum-nonconserving system. It implies that thermal transport in this system disobeys the Fourier law, in clear contrast with the existing theoretical predictions. The underlying mechanism is related to a novel effect of optical phonon-phonon interactions. These findings may open a new avenue for further exploring thermal transport in low dimensions.
\end{abstract}
\maketitle

Transport phenomena are ubiquitous in nature~\cite{Book}, such as those observed in plasmas~\cite{Plasmas}, glassy materials~\cite{Glass}, biological cells~\cite{Biology}, cold atoms~\cite{Cold}, and human mobility~\cite{Human}. Among them heat conduction is a long-standing topic. In three-dimensional bulk materials, heat conduction obeys the Fourier law, $J= - \kappa \nabla T$, i.e., the heat flux $J$ is proportional to the temperature gradient, $\nabla T$, with $\kappa$ being the thermal conductivity independent of the system size. This empirical law, together with Ohm's law for electrical conduction and Fick's law for gas diffusion, has established a general diffusive picture for various transport phenomena. Nevertheless, as pointed out by Peierls~\cite{Peierls}, unlike electrical conduction and gas diffusion, thermal transport is a challenge with many false starts, and many theories which have overlooked some essential features. At present, whether the Fourier law is valid in lower dimensions, and if yes, under what conditions, remains open to debate~\cite{Report2016}. In fact, due to the strong spatial constraints in low dimensions, anomalous non-Fourier heat transport is usually observed if the momentum is conserved in the system. With simplified microscopic models, various theoretical studies have revealed that $\kappa$ increases with the systems size~\cite{Lepri1997, Report2003, Report2008}, suggesting a scenario very similar to the problem of long-time correlations in fluids~\cite{Longtail-1, Longtail-2} or localization in disordered media~\cite{Localization}. In addition, theoretical studies have also predicted that in one-dimensional (1D) and two-dimensional (2D) cases,  anomalous thermal conduction falls into general universal categories, which has been verified by recent experimental studies of heat transport in individual single-wall nanotube~\cite{Nanotube-1, Nanotube-2} as well as in graphene~\cite{Graphene-1, Graphene-2}. These progress are of great theoretical and practical interest.

An efficient method to explore the origin of these anomalies is to focus on the system's conservation laws and study their equilibrium correlation functions~\cite{Politi2005, Zhao2006, Denisov2011, Report2015}. Early works~\cite{Zhao1998, Prosen2000, Grassberger2002, Narayan2002, LeeDadswell2005, Olla2006, Dhar2007} found that momentum conservation is a key factor, which leads to slow scattering of long-wavelength acoustic phonons~\cite{Lepri1997} and eventually results in a L\'{e}vy walk superdiffusive heat spreading in 1D case with two side peaks on the fronts~\cite{Denisov2011}. According to the recent nonlinear fluctuating hydrodynamic theory (NFHT)~\cite{Beijeren2012, Spohn2014}, a general 1D anharmonic chain without pinning has three conservation laws of, respectively, momentum, particle (stretch), and energy. This theory transforms the conserved quantities into three interacting hydrodynamic modes (two sound modes and one heat mode) and relates the heat spreading to their correlations. In particular, the correlations of these hydrodynamic modes follow a superdiffusive scaling~\cite{Denisov2011, Report2015}
\begin{equation} \label{Scaling}
t^{1/\gamma} \rho(m,t) \simeq \rho(t^{-1/\gamma}m,t),
\end{equation}
with the scaling exponent $1< \gamma <2$, different from diffusive ($\gamma=2$) and ballistic ($\gamma=1$) propagations. Here $m$ is the correlation distance between two positions of the system, $t$ is the correlation time, $\rho(m,t)$ is the corresponding correlation functions. Combining both heat (central) and sound (side) modes, the L\'{e}vy walk heat propagation function can then be understood. In this framework, if the component particles of the system are pinned by on-site potentials that breaks momentum conservation and induces the phonon-lattice interactions~\cite{Zhao1998}, a Gaussian diffusive spreading without sound modes (side peaks) will take place~\cite{Zhao1998, Xiong2017-1}. This is also the case in the chain of rotators when the conservation of stretch is destroyed~\cite{Rotator-1, Rotator-2}.

A generic scenario of two categories is thus concluded, i.e., superdiffusive L\'{e}vy walk and diffusive Gaussian propagation, with and without sound modes (side peaks), respectively, governed by the number of the conservation laws. This leads to a more in-depth question: Does there exist any heat spreading without sound modes yet be superdiffusive? Two quite recent works studying two nonacoustic systems (without sound modes) with conserved~\cite{Olla2017} and nonconserved~\cite{Satio2017} momentum gave completely opposite conclusions. In fact, in order to derive the analytical solutions, both works have introduced \emph{stochastic} dynamics to represent certain anharmonicities of the deterministic systems~\cite{Olla2006}, which makes their analysis obscure~\cite{Satio2017}; the relation between normal or anomalous transport and nonacoustic feature, momentum conservation, and stochasticity is still unclear.

In this Letter we show that a new type of superdiffusive heat spreading but without sound modes can occur in a typical pinned nonacoustic system with purely \emph{deterministic} anharmonic dynamics. This implies that in general, a momentum-nonconserving system, if bearing nonacoustic feature, does not follow normal diffusive transport. We further reveal that this anomaly is due to optical phonon-phonon interactions instead, distinct from all other known mechanisms. These findings suggest the need to develop a new theory for nonacoustic physics and may result in new understanding to transport.

Our model is an \emph{anharmonic} chain subject to \emph{harmonic} pinning (we denote it as AHsH for short) without any stochasticity. The Hamiltonian is
\begin{equation} \label{Hamiltonian}
H= \sum_{\alpha=1}^{L} p_{\alpha}^2/2 + V(r_{\alpha+1}-r_\alpha)+ U(r_\alpha),
\end{equation}
with $p_\alpha$ and $r_\alpha$ being, respectively, the momentum and the displacement (from the equilibrium position) of the $\alpha$th particle. We consider anharmonic interparticle interactions $V(\xi)=k\frac{\xi^2}{2}+ \beta \frac{\xi^4}{4}$ ($k \geq 0$, $\beta >0$) and harmonic pinning $U(\xi)=\xi^2/2$. The pinning breaks the momentum conservation of the model and therefore can significantly affect thermal transport~\cite{Prosen2005}. For example, if both $V(\xi)$ and $U(\xi)$ are harmonic, then the system exhibits ballistic transport~\cite{Prosen2005}, similar to that in harmonic chain~\cite{Lebowitz1967}, whereas for an anharmonic $U(\xi)$, regardless of $V(\xi)$, such as the well-known $\varphi^4$ lattice, the system follows normal transport instead~\cite{Zhao1998, LeeDadswell2005, Olla2006, Spohn2014, Xiong2017-1}. Nevertheless, a system with anharmonic $V(\xi)$ and harmonic $U(\xi)$ as our AHsH model has never been fully addressed~\cite{Aoki2008}, as it has always been believed to follow diffusive transport because of the breaking of momentum conservation~\cite{Zhao1998, LeeDadswell2005, Olla2006, Spohn2014, Pereverzev2003, Aoki2002}. Such a model may be of practical interest as well because the harmonic confinements are common in experimental studies such as those with cold atoms~\cite{Eli2015, Eli2016}.

\begin{figure}
\begin{centering}
\vspace{-.3cm}\includegraphics[width=8.8cm]{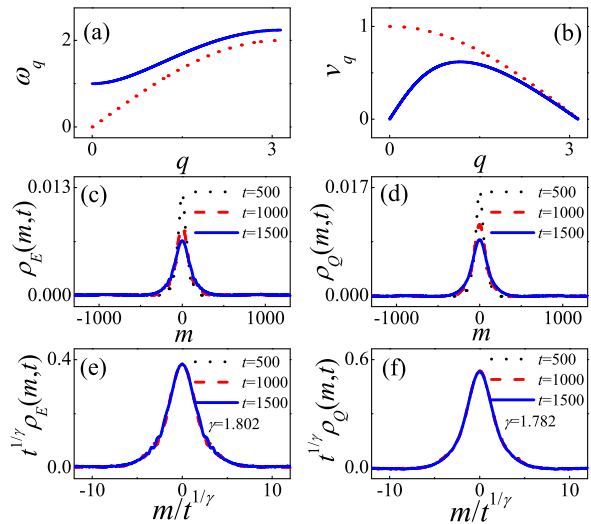} \vspace{-.9cm}
\caption{\label{Fig1} (a) The phonon dispersion relation and (b) the phonon speed of the AHsH model, where the counterpart of the harmonic chain without pinning (dotted line) is compared. (c) and (d) shows $\rho_{E}(m,t)$ and $\rho_{Q}(m,t)$ respectively for
three long times for the AHsH model with $V(\xi)= \xi^4/4$, and (e) and (f) gives the corresponding rescaled result.}
\vspace{-.3cm}
\end{centering}
\end{figure}

The AHsH model's nonacoustic feature can be seen from the phonon dispersion relation $\omega_q=[4 \sin^2(q/2)+1]^{1/2}$ and phonon speed $v_q=\frac{d \omega_q}{d q}=\frac{\sin q}{\omega_q}$ (where $q$ is the wave number) obtained from the harmonic approximation of the Hamiltonian~\eqref{Hamiltonian}. Compared with the conventional harmonic chain, $\omega_q$ now is purely optical [see Fig.~\ref{Fig1}(a)] and $v_q$ is nonmonotonic [see Fig.~\ref{Fig1}(b)]. In particular, the maximum value of $v_q$, denoted by $v_{ q*}$, takes a value of  $v_{q*}\simeq 0.618$ instead at $q^* \simeq 1.179$ and correspondingly $\omega_{q*} \simeq 1.492$, in clear contrast with that of the harmonic chain where $q^*=0$, $v_{q*}=1$, and $\omega_{q*} = 0$.

First of all, we explore the strong anharmonicity region of the AHsH model, i.e., $V(\xi)= \beta \xi^4/4$ (with $k=0$). We study the energy and the heat correlation function calculated at the equilibrium state of temperature $T=0.5$ as $\rho_{E}(m,t)=\frac{\langle \Delta E_{i+m}(t) \Delta E_{i}(0) \rangle}{\langle \Delta E_{i}(0) \Delta E_{i}(0) \rangle}$ and $\rho_{Q}(m,t)=\frac{\langle \Delta Q_{i+m}(t) \Delta Q_{i}(0) \rangle}{\langle \Delta Q_{i}(0) \Delta Q_{i}(0) \rangle}$ with $E_i(t)$ and $Q_i(t)$ being, respectively, the energy and heat density at the \emph{coarse-grained} location $i$ and time $t$  and $\Delta E_i(t)$ and $\Delta Q_i(t)$ being their fluctuations (see~\cite{Zhao2006, Denisov2011, Xiong2017-1, Zhao2013} and supplementary material~\cite{SM} for the simulation details).

The results of $\rho_E(m,t)$ and $\rho_Q(m,t)$ at three times are given in Figs.~\ref{Fig1}(c) and (d). Indeed, there are no side peaks in both of them due to the breaking of momentum conservation. But do they follow Gaussian diffusive propagation? A scaling analysis by assuming Eq.~\eqref{Scaling} indicates that $\gamma \simeq 1.802$ and $\gamma \simeq 1.782$ [see Figs.~\ref{Fig1}(e) and (f)], respectively, showing that the answer is negative. (The value of $\gamma$ is extracted from the time scaling of the central peak's height $H_c^{E, Q}$ of $\rho_{E, Q}(m,t)$ with $H_c^{E, Q} \sim t^{-1/\gamma}$; see~\cite{SM}). Such very close $\gamma$ values imply that, in the strong anharmonicity region, both $\rho_E(m,t)$ and $\rho_Q(m,t)$ of AHsH model follow the same spreading, in analogy with that observed in the momentum-nonconserving $\varphi^4$ lattice~\cite{Zhao2006, Zhao2013}, but different from the momentum-conserving Fermi-Pasta-Ulam-$\beta$ (FPU-$\beta$) model~\cite{Zhao2013}. We therefore focus on $\rho_Q(m,t)$ only in the following.

Such nondiffusive propagating is quite unexpected. According to NFHT~\cite{Spohn2014}, in momentum-nonconserving systems like the AHsH model, the only conserved field is energy and other nonconserved quantities are of no importance (their correlations will vanish), which is evident to diffusive transport. In order to clarify if this is still the case for our model, we have calculated all the $3 \times 3$ correlation functions for energy and two other nonconserved quantities (stretch and momentum) and found that the correlations of the latter decay slowly, on the contrary to the theoretical prediction (see~\cite{SM}). In what follows we will analyze the mechanism of the observed nondiffusive propagating based on this fact.

\begin{figure}
\begin{centering}
\vspace{-.3cm} \includegraphics[width=8.8cm]{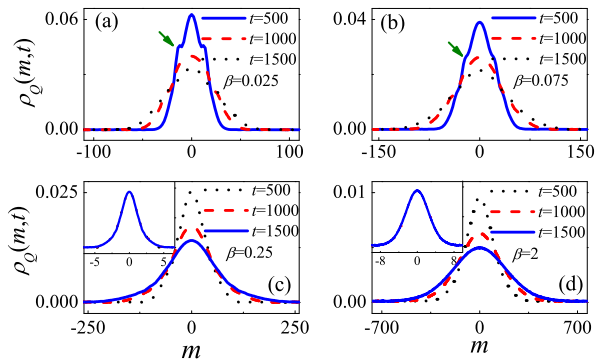} \vspace{-.8cm}
\caption{\label{Fig2} The correlation function $\rho_{Q}(m,t)$ for the AHsH model with $V(\xi)= \beta \xi^4/4$. The inset in (c) and (d) shows the rescaled results in the main panel with $\gamma=1.905$ and 1.692.} \vspace{-.3cm}
\end{centering}
\end{figure}

Before proceed we show here why the side peaks are absent but $\rho_Q(m,t)$ is yet non-Gaussian. In fact, this is similar to a superdiffusive L\'{e}vy walk density~\cite{Denisov2011} with truncated side peaks. In the original L\'{e}vy walk model~\cite{Report2015}, a particle moves after a sojourn time drawn from a probability distribution function whose tail decays in a power law, while its velocity is a finite constant $\pm v$ that leads to the side peaks. Whenever a confinement to the motion of particle is introduced, the side peaks will be truncated. This is the case in cold atoms of a Sisyphus cooling lattice with harmonic confinements~\cite{Eli2015, Eli2016}. To check if the truncation mechanism works in our model, we study the dependence of $\rho_Q(m,t)$ on $\beta$. As shown in Fig.~\ref{Fig2}, increasing anharmonicity in $V(\xi)$ would enhance the central peak and improve the scaling-invariance property, while the harmonicity in $U(\xi)$ together with this anharmonicity gradually cuts off the side peaks, but it still involves some wave-like propagations on the fronts [see Figs.~\ref{Fig2}(a) and (b), for a short time and under a small $\beta$]. This in turn affects the central peak's scaling and results in a heavy tail close to the fronts. Such heavy tails then make $\rho_Q(m,t)$ nondiffusive for a large enough $\beta$ [see Fig.~\ref{Fig2}(c)]. This effect seems not to vanish for an even stronger anharmonicity [see Fig.~\ref{Fig2}(d)].

\begin{figure}
\begin{centering}
\vspace{-.3cm} \includegraphics[width=8.8cm]{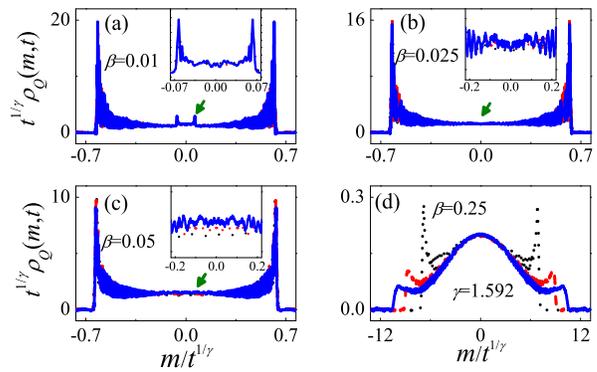} \vspace{-.9cm}
\caption{\label{Fig3} Rescaled $\rho_{Q}(m,t)$ for the AHsH model with $V(\xi)= \xi^2/2 +\beta \xi^4/4$. In (a)-(c), $\gamma=1$ and the inset shows the center part; In (d) $\gamma=1.592$. In each panel three lines are for $t=500$, 1000, and 1500 (see the legend of Fig.~\ref{Fig1}).} \vspace{-.3cm}
\end{centering}
\end{figure}

Next, let us explore the weak anharmonicity region which also helps reveal the truncation. To this end we set $k=1$ and $\beta \ll 1$ for $V(\xi)= \xi^2/2 +\beta \xi^4/4$. The results are given in Fig.~\ref{Fig3}. As for $\beta=0$ the spread has been repeatedly confirmed to be ballistic~\cite{Lebowitz1967, Prosen2005, Xiong2017-2}, we take $\gamma=1$ to rescale $\rho_Q(m,t)$ for $\beta \le 0.05$ [see Figs.~\ref{Fig3}(a)-(c)]. As shown, the predicted U shape~\cite{Xiong2017-2} of the wave-like spread can be observed and the fronts, moving with $v=v_q*$ [see Fig.~\ref{Fig1}(b)], contribute dominantly to $\rho_Q(m,t)$. This implies that optical phonons with large velocities play a major role in wave-like heat spread. As anharmonicity increases, though the wave-like propagation still keeps, the scaling may begin to deviate from the ballistic one and the central peak begins to form [see Fig.~\ref{Fig3}(c) and the inset] due to the competition between phonon dispersion and nonlinearity~\cite{Xiong2017-3}.

Note that the result at $\beta \to 0$ limit has another interesting feature [see Fig.~\ref{Fig3}(a) for $\beta=0.01$]: In addition to the whole big U shape one can identify a small one, close to center and of a small velocity $v \simeq 0.06$, nearly 10\% of $v_q*$. To our knowledge, such a profile with a double-U shape has been found only in a system with two branches of phonons~\cite{Xiong2017-2}. It might be related to the nonmonotonic feature of $v_q$ and might be found in other nonacoustic systems~\cite{Olla2017, Satio2017}.

\begin{figure}
\begin{centering}
\vspace{-.3cm}\includegraphics[width=8cm]{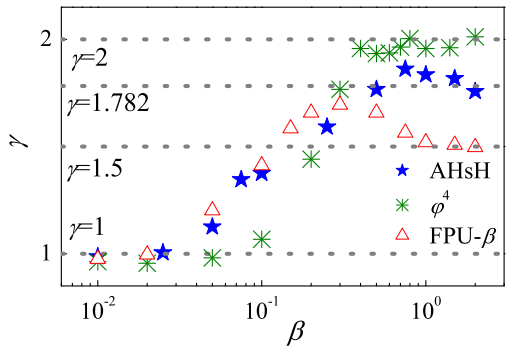} \vspace{-.8cm}
\caption{\label{Fig4} The dependence of $\gamma$ on $\beta$ for the AHsH model with $V(\xi)= \xi^2/2 +\beta \xi^4/4$, for the  $\varphi^4$ model, and for the FPU-$\beta$ model.}
\vspace{-.3cm}
\end{centering}
\end{figure}

As $\beta$ increases further, $\gamma=1$ scaling will collapse and the superdiffusive scaling of the central peak become more and more evident [see Fig.~\ref{Fig3}(d)]. In particular, the wave-like spread led by the fast optical phonons decays, signifying a truncation process. As a result, a crossover from ballistic to superdiffusive transport takes place. To capture this crossover, in Fig.~\ref{Fig4} we show how $\gamma$ depends on $\beta$ for the AHsH model. Indeed, a change from $\gamma=1$ to $1.5<\gamma<2$ is observed. This crossover is similar to that observed in the FPU-$\beta$ chain [with $V(\xi)= \xi^2/2 + \beta \xi^4/4$ and $U(\xi)=0$ in terms of Hamiltonian~(\ref{Hamiltonian})]~\cite{Xiong2016} and that in the $\varphi^4$ lattice [with $V(\xi)= \xi^2/2 $ and $U(\xi)= \xi^2/2 + \beta \xi^4/4$ instead]~\cite{Xiong2017-1}. However, despite this seeming similarity, important difference exists for large $\beta$. In the $\varphi^4$ lattice, a final diffusive spread ($\gamma=2$) will be reached. In the FPU-$\beta$ chain, eventually $\gamma$ will be between $\gamma=1.5$ and $\gamma \simeq 1.618$~\cite{Xiong2017-4, SpohnArxiv, PopkovPNAS}. In comparison, in the AHsH model $\gamma$ tends to stay at a value around $1.782$. This implies that our model follows a different sequence of universality classes, in qualitative agreement with the result in a nonacoustic system with stochastic dynamics~\cite{Satio2017}.

Now we discuss the mechanism for the observed universality sequence in the AHsH model. To this end let us turn to the momentum spreading. Recall that in the momentum-conserving FPU-$\beta$ chain, in spite of existence of phonon-phonon interactions~\cite{Lepri1997}, the long-wavelength acoustic phonons only damp very weakly. As a consequence, the momentum spreading features only two sound peaks on the fronts (see~\cite{SM}). However, for momentum-nonconserving systems like the $\varphi^4$ lattice, there are no acoustic phonons, thus the front peaks disappear. Indeed, phonons here are scattered by anharmonic pinning, an effect of phonon-lattice interactions~\cite{Zhao1998}. For large enough anharmonicity, when this effect is dominant, the momentum correlation will disappear~\cite{Zhao2013}, consistent with the conjecture of NFHT~\cite{Spohn2014}. Hence, momentum spreading helps distinguish possible effects of phonon-phonon and phonon-lattice interactions, and here we adopt it to probe the AHsH model.

\begin{figure}
\begin{centering}
\vspace{-.3cm} \includegraphics[width=8.8cm]{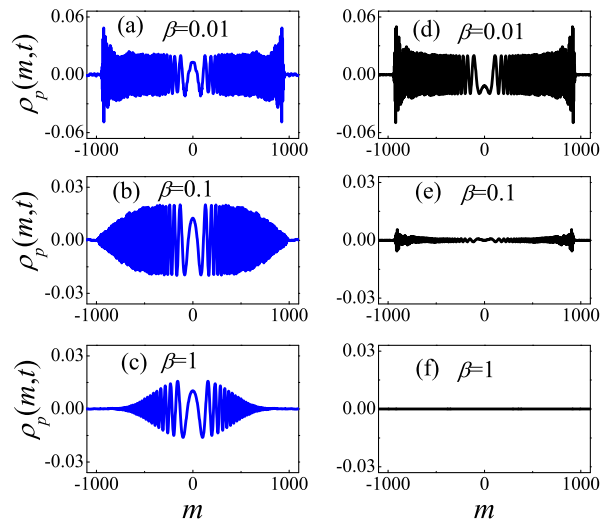} \vspace{-.9cm}
\caption{\label{Fig5} Comparison of the momentum correlation function $\rho_{p}(m,t)$ at $t=1500$ between the AHsh model with $V(\xi)= \xi^2/2 +\beta \xi^4/4$ (a)-(c) and the $\varphi^4$ lattice (d)-(f).}
\vspace{-.3cm}
\end{centering}
\end{figure}

In Fig~\ref{Fig5}, the momentum correlation function $\rho_{p}(m,t)=\frac{\langle \Delta p_{i+m}(t) \Delta p_{i}(0) \rangle}{\langle \Delta p_{i}(0) \Delta p_{i}(0) \rangle}$ for both the AHsH and the $\varphi^4$ model are compared. Note that anharmonicity is provided by $V(\xi)$ [$U(\xi)$] only in the former (latter). Suppose that if all optical phonons are completely damped, $\rho_{p}(m,t)$ will vanish, which is indeed the case in the $\varphi^4$ lattice with a strong anharmonicity. With this in mind, Fig.~\ref{Fig5} shows a great distinction: Although for a small $\beta$, both models' $\rho_{p}(m,t)$ are similar [see Figs.~\ref{Fig5}(a) and (d)], suggesting ballistic wave-like spreading where fast phonons contribute more [see also Figs.~\ref{Fig3}(a) to (c)], as $\beta$ increases, while in the $\varphi^4$ lattice all the fast and slow phonons decay increasingly rapidly [see Figs.~\ref{Fig5}(e) and (f)], which implies that the attenuation of phonons is insensitive to their velocities, in the AHsH model fast phonons decay quickly but slow phonons decay slowly, resulting in a spindle wave pattern [see Fig.~\ref{Fig5}(b)], and this pattern remains for an even stronger anharmonicity [see Fig.~\ref{Fig5}(c)].

Such a great distinction implies that, for momentum-nonconserving systems, the propagation property is sensitive to whether the anharmonicity is in $U(\xi)$ or $V(\xi)$. Because of this, in the $\varphi^4$ lattice, the phonon-lattice interactions play a major role, while in the AHsH model, a different mechanism of optical phonon-phonon interactions takes effect. This mechanism is also distinct from the known effect of acoustic phonon-phonon interactions as in the momentum-conserving FPU-$\beta$ chain where a different momentum spreading behavior (see~\cite{SM}) follows.

To summarize, we have shown that a new type of anomalous heat spreading without sound modes peaks yet be superdiffusive, can occur in a \emph{momentum-nonconserving}, \emph{nonacoustic} AHsH model with \emph{deterministic} dynamics. It is found that this propagation can be described by a truncated L\'{e}vy walk model resulted from the unexpected correlations of the nonconserved fields. Such a non-Gaussian spreading property is quite surprising, because for anharmonic momentum-nonconserving systems, Gaussian heat spreading has always been expected based on the existing theories. It implies that the current theoretical models, such as NFHT, are inadequate for describing the nonacoustic AHsH system, and therefore a further developed theory is required~\cite{Satio2017}. In developing the theory, as our study suggests, whether the anharmonicity is in the interparticle potential or in the pinning potential should be dealt with carefully.

We have further revealed that anharmonicity in the interparticle potential only leads to a special property of optical phonon-phonon interactions. This is in contrast with that of phonon-lattice interactions~\cite{Zhao1998} in the momentum-nonconserving $\varphi^4$ lattice and that of the acoustic phonon-phonon interactions~\cite{Lepri1997} in the momentum-conserving FPU-$\beta$ chain. Combining all these observations, it can be concluded that for deterministic nonacoustic systems, the phonon-phonon interactions, even of optical type, exclusively gives rise to superdiffusive transport.

Our study may shed new light on the thermal transport problem. It may be interesting to experimental studies and applications as well, because in nanosystems~\cite{Nanotube-1,Nanotube-2,Graphene-1,Graphene-2}, the finite size effects and pinnings, anharmonic or harmonic as an approximation, can not be avoided in many cases~\cite{Eli2015,Eli2016}.

\begin{acknowledgments}
We are indebted to Jiao Wang for his kind help in preparing the revised manuscript. We also thank R. Livi, S. Lepri, H. Spohn, and S. V. Dmitriev for valuable comments. D. X. is supported by NNSF (Grant No. 11575046) of China, NSF (Grant No. 2017J06002) of Fujian province, the training plan for distinguished young researchers of Fujian provincial department of education, and the Qishan scholar research fund of Fuzhou university. Y. Z. is supported by NNSF (Grant No. 11335006) of China and the President Foundation (Grant No. 20720150036) of Xiamen University.
\end{acknowledgments}

\pagebreak
\clearpage
\begin{center}
\textbf{\large \large Supplementary Material for
`One-dimensional Superdiffusive Heat Propagation Induced by Optical Phonon-Phonon Interactions'}
\end{center}
\setcounter{equation}{0} \setcounter{figure}{0}
\setcounter{table}{0} \setcounter{page}{1} \makeatletter
\renewcommand{\theequation}{S\arabic{equation}}
\renewcommand{\thefigure}{S\arabic{figure}}
\renewcommand{\bibnumfmt}[1]{[S#1]}
\renewcommand{\citenumfont}[1]{S#1}
\emph{1. Simulation of the correlation functions.}---We study the following three correlation functions:
\begin{equation}
\rho_{Q}(m,t)=\frac{\langle \Delta Q_{i+m}(t) \Delta Q_{i}(0) \rangle}{\langle \Delta Q_{i}(0) \Delta Q_{i}(0) \rangle},
\end{equation}
\begin{equation}
\rho_{E}(m,t)=\frac{\langle \Delta E_{i+m}(t) \Delta E_{i}(0) \rangle}{\langle \Delta E_{i}(0) \Delta E_{i}(0) \rangle},
\end{equation}
and
\begin{equation}
\rho_{p}(m,t)=\frac{\langle \Delta p_{i+m}(t) \Delta p_{i}(0) \rangle}{\langle \Delta p_{i}(0) \Delta p_{i}(0) \rangle},
\end{equation}
where $\langle \cdot \rangle$ represents the spatiotemporal average; $\triangle \chi=\chi-\langle \chi \rangle$ is the fluctuations of the corresponding quantity $\chi$, specifically, the heat energy $Q$, the total energy $E$, and the momentum $p$; $i$ and $t$ are the space and time variables, and $m$ gives the correlation distance.

To numerically simulate the correlation functions, we follow the method adopted in Ref.~\cite{S_Zhao2013}. One key point is that $i$ here is a \emph{coarse-grained} space variable but not the particle's index. So, in practice one has to divide the chain into several bins of equal size and define all the relevant quantities in these bins. The $i$th bin's quantity $\chi_i(t)$ (at time $t$) then is calculated by a summation of the corresponding quantity of the relevant particles within the bin. Under this description, $p_i(t)$ and $E_i(t)$ are easily computed since their expressions for each particle are implicitly known. According to the conventional hydrodynamic theory~\cite{S_Forster,S_Liquid,S_Beijeren2012}, the $i$th bin's heat energy follows
$Q_i(t) \equiv E_i(t)-\frac{(\langle E \rangle + \langle F\rangle) g_i(t)}{\langle g \rangle}$,
where $g_i$ is the number of particles within the $i$th bin (which is also easily computed) and $F_i$ is the pressure exerted on the bin. Now the only unknown quantity is the averaged pressure $\langle F \rangle $. Fortunately, as a standard setup, we set both the equilibrium distance between particles as well as the lattice constant to unity. This makes the number of particles equal to the system size $L$, and therefore, for the symmetric interparticle potentials like $V(\xi)=k\frac{\xi^2}{2}+ \beta \frac{\xi^4}{4}$ considered here, $\langle F \rangle \equiv 0$.

For all the simulations, we first prepare a canonical equilibrium system of temperature $T=0.5$ by the Langevin heat baths~\cite{S_Lepri_Report,S_Dhar_Report}. We consider the chain with $L=4001$ particles of periodic boundary conditions, which ensures that an initial fluctuation located at the center can spread out a time up to $t=1500$. To calculate the correlation functions, we always fix the number of bins as $(L-1)/2$, which means that the bin size is $2$; but we have checked and verified that for a larger bin size, the simulation results remain the same. We employ the Runge-Kutta algorithm of 7th to 8th order with a time step of $h=0.05$ to evolve the system. Each equilibrium system is prepared by evolving the system for a long enough time ($>10^7$ time units) from properly assigned initial random states, then the system is evolved in isolation. Finally, we use ensembles of size about $8\times10^9$ to compute the correlation functions.
\begin{figure}
\begin{centering}
\vspace{-0.5cm}\includegraphics[width=7cm]{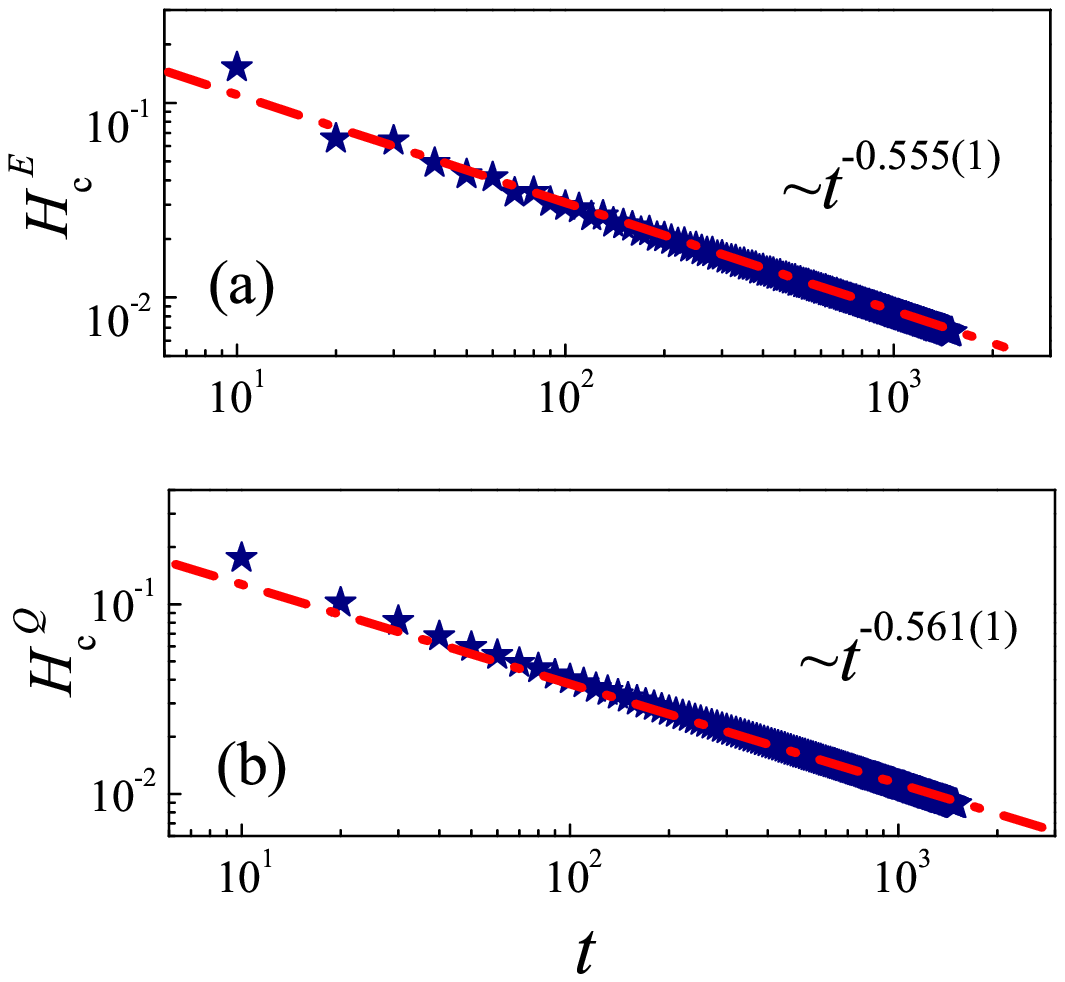} \vspace{-.8cm}
\caption{\label{SFig1} The height $H_c^{E}$ ($H_c^{Q}$) of the central peak of $\rho_{E}(m,t)$ [$\rho_{Q}(m,t)$] vs $t$.}
\vspace{-.6cm}
\end{centering}
\end{figure}

\emph{2. Method to extract $\gamma$.}---To extract the scaling exponent $\gamma$, we explore the time scaling of the energy and the heat correlation functions' central peak. Fig.~\ref{SFig1} presents the height $H_c^{E,Q}$ of the central peak of $\rho_{E,Q}(m,t)$ (in Fig.~1) with $t$. Both show good scalings: $\sim t^{-1/\gamma}$ with $\gamma=\frac{1}{0.555} \simeq 1.802$ and $\gamma =\frac{1}{0.561} \simeq 1.782$, respectively.  These $\gamma$ values are used to rescale the corresponding correlation functions in Figs.~1(e,f).
\begin{figure*}
\begin{centering}
\vspace{-0.5cm} \includegraphics[width=12cm]{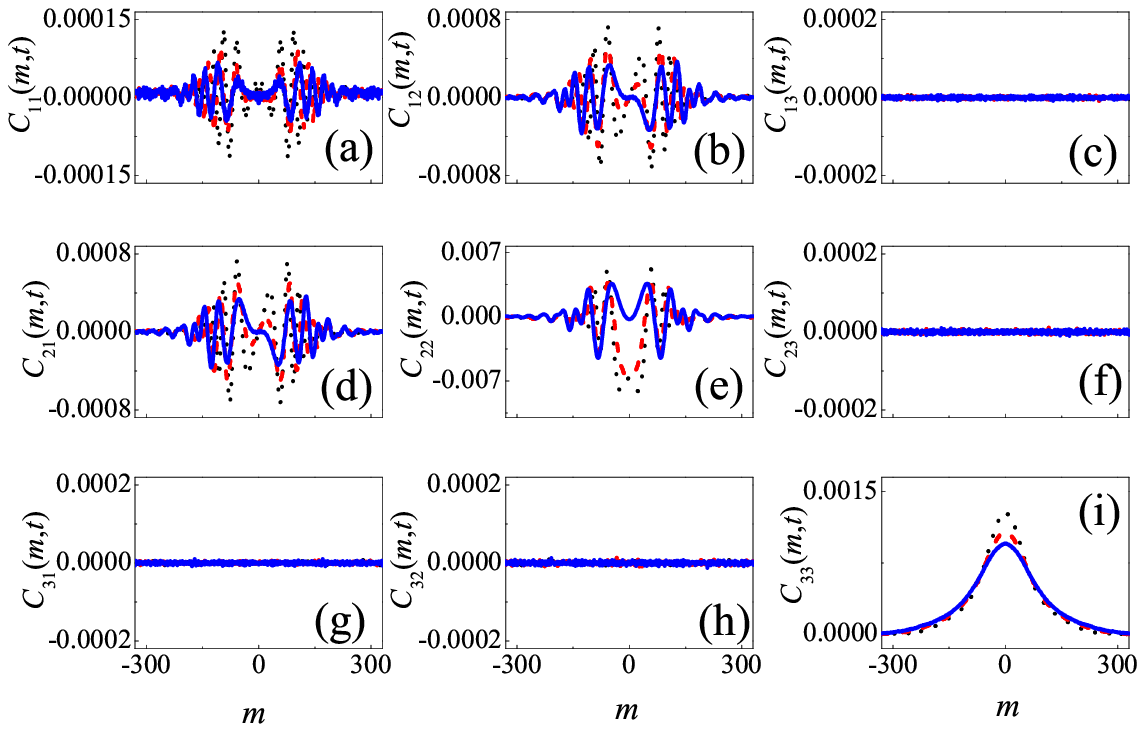} \vspace{-0.7cm}
\caption{\label{SFig2} The full $3 \times 3$ correlation functions for the stretch, the momentum, and the energy for the AHsH model with $V(\xi)= \xi^4/4$, for three long times: $t=600$ (dotted), $t=800$ (dashed), and $t=1000$ (solid).} \vspace{-0.7cm}
\end{centering}
\end{figure*}

\emph{3. Correlation functions under the NFHT scheme.}---According to NFHT~\cite{S_Spohn2014}, in momentum-nonconserving systems like the AHsH model, energy is the only conserved field and other nonconserved quantities (stretch and momentum) are of no importance (their correlations will vanish), which is evident to diffusive heat transport. To clarify if this is still the case in our AHsH model, we here calculate the full $3 \times 3$ correlations for the stretch $s_m=r_{m+1}-r_m$, the momentum $p_m$, and the energy $E_m$. Note that in the conventional hydrodynamic theory~\cite{S_Forster,S_Liquid,S_Beijeren2012}, another local field of the particle'a number, instead of the stretch, is considered, but this will not make a qualitative difference of correlations~\cite{S_Xiong2017-4}.

In the framework of NFHT, we consider the correlation functions of elements:
\begin{equation} \label{CC}
C_{\mu\nu} (m,t)=\langle u_{\mu}(m,t) u_{\nu}(0,0) \rangle,
\end{equation}
where $u_1$, $u_2$, and $u_3$ are, respectively, the corresponding fields' small fluctuations, defined by $u_1(m,t)=s_m(t)-\langle s \rangle$, $u_2(m,t)=p_m(t)-\langle p \rangle$, and $u_3(m,t)=E_m(t)-\langle E \rangle$; $\mu, \nu \equiv 1, 2, 3$. These correlations are presented in Fig.~\ref{SFig2}. Now it can be concluded that, although here energy is the only conserved quantity, it only implies no correlations between energy and other two nonconserved fields [see $C_{13}(m,t)$, $C_{23}(m,t)$, $C_{31}(m,t)$, and $C_{32}(m,t)$ in Figs.~\ref{SFig2}(c,f,e,h)], whereas there are still correlations between stretch and momentum themselves [see $C_{11}(m,t)$, $C_{12}(m,t)$, $C_{21}(m,t)$, and $C_{22}(m,t)$ in Figs.~\ref{SFig2}(a,b,d,e)]. More-importantly, as time increases these correlations decay very slowly, on the contrary to the theoretical prediction.
\begin{figure}
\begin{centering}
\vspace{-.5cm} \includegraphics[width=7cm]{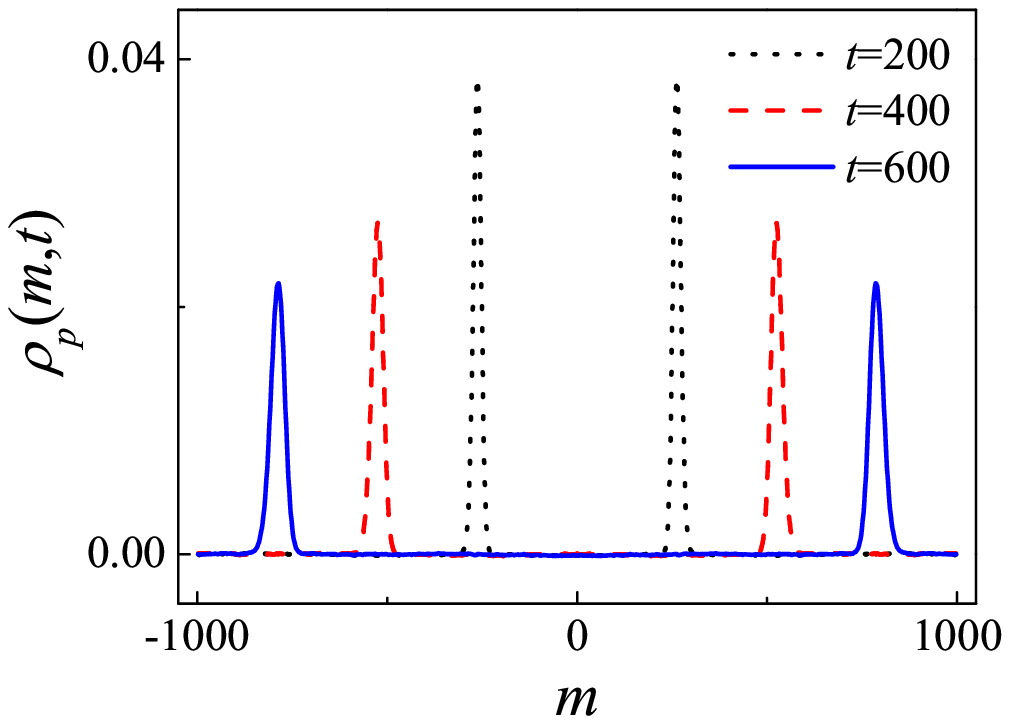} \vspace{-.8cm}
\caption{\label{SFig3} The momentum correlation function $\rho_p(m,t)$ in the FPU-$\beta$ chain ($\beta=1$).} \vspace{-.6cm}
\end{centering}
\end{figure}

\emph{4. Momentum correlation in the FPU-$\beta$ chain.}---For the momentum-conserving FPU-$\beta$ chain, in spite of existence of acoustic phonon-phonon interactions, the long-wavelength acoustic phonons only damp very weakly. As a result, the momentum correlation function features only two sound peaks on the fronts. This is a known result that has been addressed in the literature~\cite{S_Lepri1998}. Here, we present a typical result in Fig.~\ref{SFig3} for reference.

\end{document}